\documentclass[twocolumn]{aastex63}

\usepackage{amsmath}
\usepackage{graphicx}
\usepackage{tabularx}
\usepackage{dcolumn}
\usepackage{bm}
\usepackage{xcolor}

\newcommand{\dHybridR}{{\it dHybridR}}
\newcommand{\ocii}{\Omega_{ci}^{-1}}
\newcommand{\di}{d_{i}}
\newcommand{\va}{v_{A0}}

\newcommand{\vsh}{v_{\rm sh}}
\newcommand{\rt}{R_{\rm tot}}
\newcommand{\trt}{\Tilde{R}_{\rm tot}}
\newcommand{\trs}{\Tilde{R}_{\rm sub}}
\newcommand{\rs}{R_{\rm sub}}
\newcommand{\pinj}{p_{\rm inj}}
\newcommand{\w}[1]{v_{A,#1}}
\newcommand{\tu}[1]{\Tilde{u}_{#1}}

\submitjournal{ApJ}

\shorttitle{CR-Modified Shocks II: Particle Spectra}
\shortauthors{Caprioli et al.}

\graphicspath{{./}{figures/}}

\begin{document}

\title{Kinetic Simulations of Cosmic-Ray–Modified Shocks II: Particle Spectra}

\correspondingauthor{Damiano Caprioli}
\email{caprioli@uchicago.edu}

\author[0000-0003-0939-8775]{Damiano Caprioli}
\author[0000-0002-2160-7288]{Colby C. Haggerty}
\affiliation{Department of Astronomy and Astrophysics, University of Chicago, 5640 S Ellis Ave, Chicago, IL 60637, USA}
\author[0000-0003-2480-599X]{Pasquale Blasi}
\affiliation{Gran Sasso Science Institute, Viale F.\ Crispi 7, 67100 L’Aquila, Italy}
\affiliation{INFN/Laboratori Nazionali del Gran Sasso, Via G. Acitelli 22, Assergi (AQ), Italy}

\date{\today}

\begin{abstract}
Diffusive shock acceleration is a prominent mechanism for producing energetic particles in space and in astrophysical systems. 
Such energetic particles have long been predicted to affect the hydrodynamic structure of the shock, in turn leading to CR spectra flatter than the test-particle prediction. 
However, in this work along with a companion paper, \cite{haggerty+20}, \url{https://arxiv.org/abs/2008.12308}, we use self-consistent hybrid (kinetic ions-fluid electrons) simulations to show for the first time how CR-modified shocks actually produce {\it steeper} spectra.
The steepening is driven by the enhanced advection of CRs embedded in magnetic turbulence downstream of the shock, in what we call the  ``postcursor''. 
These results are consistent with multi-wavelength observations of supernovae and supernova remnants and have significant phenomenological implications for space/astrophysical shocks in general.
\end{abstract}

\section{Introduction} \label{sec:intro}
Diffusive Shock Acceleration \citep[DSA,][]{krymskii77,bell78a,blandford+78,axford+78} is a ubiquitous mechanism for producing relativistic particles and non-thermal emission in many astrophysical environments.
This special case of first-order Fermi acceleration involves particles diffusing back and forth across the shock discontinuity and is particularly appealing because it produces power-law distributions in momentum space that depend solely on the shock compression ratio, $r=\rho_2/\rho_1$,  i.e., the ratio of  downstream to upstream plasma density. 
For a gas with adiabatic index $\gamma$, the compression ratio $r$ and the corresponding momentum slope $q_{\rm DSA}$ read
\begin{equation}\label{eq:qdsa}
    r=\frac{\gamma+1}{\gamma-1+2/M^2},     
    \quad
    q_{\rm DSA}= \frac{3r}{r-1};
\end{equation}
for strong shocks with Mach number $M\gg1$ and $\gamma=5/3$, $r\to 4$, this corresponds to the universal DSA slope of $q_{\rm DSA}\to 4$ and hence relativistic energy spectra $\propto E^{-2}$.

When accelerated particles (henceforth cosmic rays, CRs) carry a non-negligible fraction of the shock momentum/energy, they can no longer be regarded as test-particles and CR-modified shocks arise, where both the shock dynamics and particle spectra deviate from the standard predictions \citep[e.g.,][]{drury-volk81a,drury83,jones+91,malkov+01}. 
The standard theory of non-linear DSA (NLDSA) predicts that the CR pressure produces an upstream precursor, in which the incoming fluid is slowed down, compressed, and heated;
CR-driven currents in the upstream also trigger the generation of strong magnetic turbulence that, in addition to fostering CR scattering, may also have a dynamical role \citep{vladimirov+06,caprioli+08,caprioli+09a}.
Throughout the paper we will indicate with subscript 0,1, and 2 quantities measured at upstream infinity, immediately upstream of the shock, and downstream, respectively.
A general feature of NLDSA is that the precursor weakens any strong shock into a subshock with compression ratio $\rs\equiv \rho_2/\rho_1\lesssim 4$, while the overall compression ratio increases to $\rt\equiv \rho_2/\rho_0\gtrsim 4$. 

NLDSA effects lead to spectra modulated based on the effective compression ratio felt by particles with a given momentum:
low-energy CRs that remain confined closer to the subshock probe $\rs$, while the largest-energy CRs experience $\rt$, eventually resulting in concave spectra, steeper/flatter than $p^{-4}$ at low/high momenta.
The boundary between these regimes is at the lowest momentum where CRs carry a non-negligible pressure, typically at trans-relativistic energies;
therefore, all of the non-thermal emission in astrophysical environments should be determined by relativistic CRs with a spectrum \emph{flatter} than $p^{-4}$; for instance, $q\simeq 3.5$ for $\rt=7$.

\subsection{The Clash Between Theory and Observations}
The discrepancy between the NLDSA theory and the phenomenology of strong shocks became compelling with the first GeV observations of Galactic supernova remnants (SNRs) that ---coupled with pre-existing TeV observations--- did not confirm the existence of concave spectra.
Even more so, such observations strongly hinted that efficient CR acceleration must coexist with spectra steeper than $p^{-4}$  \citep{caprioli11}.

The case for steep spectra was already raised by radio observations of extra-galactic SNe \citep[e.g.,][]{chevalier+06}, which are typically fitted with spectra as steep as $p^{-5}$.
However, the relativistic electrons responsible for such a synchrotron emission are likely sub-GeV (bearing the uncertainty in the magnetic field) and so they may potentially be accounted for by the steep part of a concave NLDSA spectrum  \citep[e.g.,][]{ellison+91,ellison+00,tatischeff09}.

Another piece of evidence that DSA should produce spectra steeper than $p^{-4}$ is linked to the origin of Galactic CRs. 
Traditional arguments based on the amplitude of the observed anisotropy favor a Galactic residence time that scales as $\sim E^{-0.3}$ \citep{blasi+11a,blasi+11b}, thereby implying that the injection spectrum be stepeer than $E^{-2}$. More recent measurements of the secondary-to-primary ratios and of the spectral shape of primary elements strongly point toward a diffusion coefficient with a non trivial structure, with a break at $\sim 300$ GV rigidity. 
For higher rigidity, the energy dependence that best describes data is again $\sim E^{0.3}$ \citep{ams16b}. The general fit to the overall spectrum requires an injection spectrum of primary CRs which is $\sim E^{-2.3}$ to $E^{-2.4}$ \citep{evoli+19a,evoli+19b}, definitely steeper than expected from standard DSA and even more so if compared with the results of NLDSA.

An extended discussion of these observational constraints can be found in \cite{caprioli15p} and \cite{caprioli+19p}. 

\subsection{Attempts to Revise DSA Theory}
Possible solutions to this discrepancy between NLDSA and observations have been put forward \citep[see][for a review]{caprioli15p}, but each of them has either a limited range of applicability or hinges on some unverified assumption about the complex CR-magnetic field interplay. 
In essence, the universal $p^{-4}$ momentum spectrum arises under the assumption that CRs are isotropized in the fluid rest frame both upstream and downstream, such that at each shock crossing they gain momentum due to a compression $\Delta u/v$, where $\Delta u=u_1-u_2$ is the difference between the upstream and downstream fluid speeds in the shock frame and $v$ is the particle speed;
the balance between such a gain and the probability  of being advected away downstream ($\propto u_2/v$) leads to power-law spectra with the canonical slope \citep{bell78a}. 
The first-order correction to isotropy is given by the diffusive flux, which is $\mathcal{O}(u/v)$ and controls the rate of acceleration, but not the slope \citep[e.g.,][]{drury83,blasi+07}.

Deviations from the universal spectrum may arise if one (or more) of the hypotheses above is (are) violated; 
here we present a brief critical summary of the main ideas suggested in the literature.

1) CRs may be isotropized in a frame that moves with the magnetic waves\footnote{Technically, magnetic fluctuations are not necessarily linear, but we will refer to them as ``waves'' for brevity.}, such that the velocity that matters is $\Tilde{u}\equiv u + v_w$, where $v_w$ is the local wave velocity, typically of the order of the Alfv\'en speed, $v_A$.
For this effect to lead to systematically steeper spectra, the correction $v_w$ has to be non-negligible with respect to $u$ and the \emph{sign} of the wave velocity both upstream and downstream has to be consistent. 
While it can be argued that in the upstream, self-generated waves travel against the CR gradient, i.e., against the flow, downstream waves are usually taken to have no preferential direction and hence $v_{w,2}\to 0$; 
with these assumptions, reasonable but never validated by kinetic simulations,
\begin{equation}
    \Tilde{r}\equiv \frac{\tu1}{\tu2}\simeq \frac{u_1-\w1}{u_2}\lesssim r
\end{equation}
and CR spectra are systematically steeper than the DSA prediction  \citep[e.g.,][]{zirakashvili+08b,caprioli+10b,caprioli12,kang+13,slane+14,kang+18}.

2) CR transport may be intrinsically anisotropic and/or inhomogeneous, which may happen for very fast ($\vsh\gtrsim 10^4$\,km\,s$^{-1}$), very oblique shocks \citep[e.g.,][]{kirk+96, bell+11}.
This scenario, which implicitly assumes that ion injection may spontaneously occur at quasi-perpendicular shocks  \citep[not granted in the absence of energetic seeds, see][]{caprioli+15,caprioli+17,caprioli+18}, may be viable for fast radio SNe, but not for middle-age/old $\gamma$-ray bright SNRs.

3) The shock dynamics may be modified by the presence of neutral hydrogen, which is coupled to the ionized component via ionization and charge-exchange processes \citep{blasi+12a,morlino+12b,morlino+13}. 
The backstreaming of neutrals that underwent charge-exchange downstream leads to the formation of a strong upstream precursor that makes the shock significantly weaker for CRs whose diffusion length is smaller than the charge-exchange mean free path \citep{blasi+12a}.
Since for $\vsh\gtrsim 3000$\,km\,s$^{-1}$ ionization dominates over charge exchange and the neutral return flux vanishes, this explanation is not viable for fast radio SNe, but works very nicely for older remnants, such as Tycho \citep{morlino+16}.
Also, strong ion-neutral damping might steepen the spectrum because of the partial evanescence of Alfv\'en waves  \citep{malkov+12}, but the exact amount of steepening is hard to quantify and may arguably morph into a hard cutoff above a few GeV, effectively killing the whole acceleration process.

4) Steep spectra observed in SNRs have been suggested to be due to the time convolution of spectra with different cut-offs produced in increasingly larger accelerating regions, namely where the shock is quasi-parallel to the large-scale magnetic field \citep{malkov+19}.
This effect, however, may apply in a different way to ions and electrons and in general cannot account for steep spectra in systems where the coherence length of the background magnetic field is smaller than the SNR size, e.g., in Tycho \citep{morlino+12}.
Alternatively, it has been suggested that a shock that encompasses both quasi-parallel and quasi-perpendicular regions may exhibit global spectra steeper than those found at a quasi-parallel shock \citep{hanusch+19}. 
This effect is hard to reckon with since oblique patches always appear at initially quasi-parallel shocks due to magnetic field amplification and because \cite{hanusch+19} report that oblique regions show spectra \emph{flatter} than quasi-parallel ones (see their Figure 5 of \cite{hanusch+19}). However, this effect is limited to a steepening of $\Delta q\sim 0.1$.
This is also consistent with the findings of \cite{caprioli+18}, who do not report appreciable differences in the spectra produced at parallel and oblique shocks in the presence of energetic seeds.

5) CRs may effectively loose energy upstream at the expense of the generation of magnetic turbulence, such that the energy gain from DSA per cycle is reduced \citep{bell+19}.
This possibility has been suggested very recently in an analytical framework, but never validated with self-consistent plasma simulations.

\subsection{Simulations of CR-modified Shocks}
The goal of this paper is to put forward the first evidence of the production of spectra steeper than the canonical DSA prediction in self-consistent kinetic simulations of collisionless shocks.
In this work and a companion paper (\cite{haggerty+20}, henceforth Paper I), we have used the hybrid code \dHybridR~\citep{haggerty+19a}, a particle-in-cell electromagnetic
code with kinetic ions and fluid electrons, to model the formation and long-term evolution of CR-modified shocks \emph{ab initio}. 
\dHybridR~introduces the full relativistic dynamics of ions into the Newtonian code {\it dHybrid} \citep{gargate+07}.
In  Paper I we have quantified the effects of CR production on the shock hydrodynamics, finding that, when acceleration is efficient, the shock develops the canonical precursor upstream, but also a \emph{postcursor} downstream, in which self-generated magnetic fluctuations induce a relative drift between the background plasma and the CR population.
Strong, quasi-parallel shocks, which spontaneously inject thermal particles into DSA, can channel $\sim 10-15\%$ of the shock ram pressure into CRs and $\sim 3-8\%$ into magnetic pressure in the downstream \citep{caprioli+14a,caprioli+14b, haggerty+20}. 
As a consequence of the magnetic drift in the postcursor, the non-thermal energy in CRs and magnetic field is transported away from the shock at a faster rate than in gaseous shocks (section 4.1 and 4.2 of Paper I) and the total compression ratio $\rt$ increases beyond the nominal value of $4$, up to values of $\rt\gtrsim 6$; $\rs$, instead, does not appreciably deviate from $\sim 4$.

In general, three different effects may lead to a $\rt\gtrsim 4$: 
1) the change in the effective adiabatic index of the gas+CR+magnetic field system \citep[e.g.,][]{drury83,jones+91};
2) the escape of CRs from upstream, which make the shock behave as partially radiative \citep[e.g.,][]{drury-volk81a,caprioli+09b};
3) the CR+magnetic field drift with respect to the thermal plasma in the postcursor, which also mimics a non-standard energy escape (Paper I).
In Paper I we singled out, for the first time, the prominence of the postcursor in the dynamics of CR-modified shocks over the other two effects;
we expect this to be the main effect in astrophysical environments because in strong shocks it is quantitatively dominant over the first two effects\footnote{Note that for moderate acceleration efficiencies of $10-20\%$, CRs escaping upstream of the shock carry away $\lesssim 1\%$ of the bulk energy \citep{caprioli+09b}; if spectra are steeper than $p^{-4}$, the escaping highest-energy particles take away a negligible amount of energy.}

\section{A Revised DSA Theory}
At face value, if CRs see a compression ratio $\rt>4$ their spectrum should become flatter than the test-particle prediction, but in this section we show how the postcursor introduces a fundamental correction to the classical DSA theory.
As discussed in Paper I, in the postcursor, CRs and magnetic structures travel downstream with respect to the thermal plasma with a drift speed comparable to the local Alfv\'en speed, such that:
\begin{equation}\label{eq:trt}
    \trt\simeq \frac{u_0}{u_2+\w2}\simeq \frac{\rt}{1 +\alpha}; \quad 
    \alpha\equiv \frac{\w2}{u_2},
\end{equation}
where in the numerator we set $\w0\approx 0$ because upstream infinity waves do not have a preferential direction.
The $\alpha$ parameter quantifies the effect of the postcursor-induced spectral modification; 
since $\alpha>0$, the compression ratio felt by the CRs is always smaller than the fluid one.
At the same time, for low-energy CRs that probe the subshock only
\begin{equation}
    \trs\simeq \frac{u_1-v_{A,1}}{u_2+v_{A,2}}\simeq \rs\frac{1-\alpha_1}{1 +\alpha}; \quad 
    \alpha_1\equiv \frac{\w1}{u_1}.
\end{equation}
Note that, when the magnetic field is compressed at the shock and $B_2\approx\rs B_1$, we have $\alpha= \rs^{3/2}  \alpha_1\lesssim 8\alpha_1$ and the correction due to the postcursor dominates over the one in the precursor, which was the only effect accounted for by the previous literature \citep{zirakashvili+08b,caprioli12}.
This is why we focus on the parameter $\alpha$, 
which technically might be labelled as $\alpha_2$, rather than on $\alpha_1$.

While the effect of the postcursor should steepen spectra, it is not expected to make the spectra arbitrarily steep. This is because the enhanced postcursor magnetic field originates from instabilities upstream of the shock which are in turn driven by CRs.
If the postcursor backreaction becomes too prominent, then CR spectra will become too steep, and in turn the upstream CR current will abate, thereby limiting the role of the postcursor;
in this way, the CR spectra and the postcursor should reach an equilibrium state where the spectral steepening is self-regulated.
Assessing the exact value of the steepening at equilibrium is nontrivial and requires a self-consistent calculation;
physically speaking, though, we expect self-regulation to kick in when $q\gtrsim 4$ and to become more and more effective when $q$ increases, likely saturating when $q\lesssim 5$. 

In summary, regardless of CRs feeling the total or the subshock compression ratio, the postcursor drift reduces the effective compression ratio felt by about $1+\alpha$.
In the benchmark simulation examined in Paper I, where $\alpha\approx 0.6$ and $\rt\approx 6$, this would correspond to $\trt\lesssim 4$ and hence to a spectrum \emph{steeper} than the standard $p^{-4}$ power law.

\section{Spectra in CR-modified Shocks}
\begin{figure}[t]
\includegraphics[width=0.48\textwidth, clip=true,trim= 0 0 0 0]{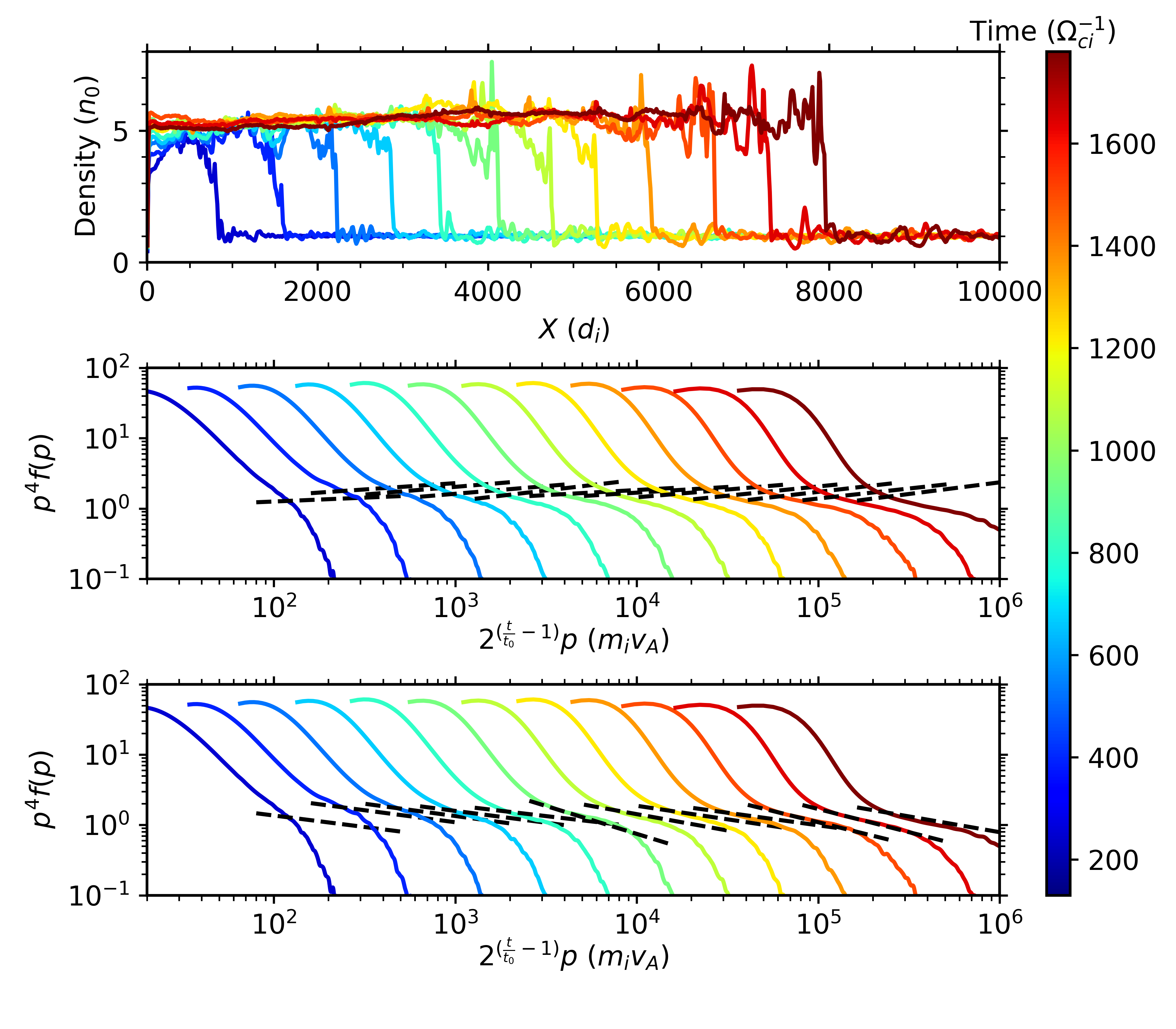}
\caption{Time evolution (color coded) of density profile and post-shock particle spectra (top and bottom two panels, respectively) for our benchmark run of a parallel shocks with $M=20$.
Despite the fluid shock compression ratio $\rt\to 6$, the CR non-thermal tail is significantly steeper than the standard prediction (Equation \ref{eq:qdsa}, middle panel) and in good agreement with the revised prescription (Equation \ref{eq:qtilde}, bottom panel). 
}\label{fig:evolution}
\end{figure}
In Paper I, we presented the results from self-consistent hybrid simulations of quasi-parallel shocks that are efficiently accelerating particles and generating magnetic fields.
We refer to that paper for the details of the simulation setup;
here, we recall that density $\rho$ and magnetic field $B$ are normalized to their far upstream values, speeds to the Alv\'en speed $\va \equiv B_0/\sqrt{4\pi \rho_0}$, lengths to the ion inertial length $\di$, and time to the inverse ion cyclotron frequency $\ocii$.
Since shocks are produced using a reflecting wall, the downstream is at rest in the simulations and, in this frame, the upstream has a speed $\vsh = M\va$, where $M$ is the Alfv\'enic Mach number (comparable to the sonic Mach number).

Our benchmark run focuses on the long-term evolution of a parallel shock with $M=20$, in which CR-induced modifications are evident.
The top panel of Figure \ref{fig:evolution} shows how after a few hundred $\ocii$, $\rt$ increases beyond the canonical value of $4$ and $\rt\to 6$ at later times.
The standard prediction is that the CR momentum spectrum should flatten with time: such expected spectra with $q(r)\equiv 3\rt/(\rt-1)$ are shown with dashed lines in the middle panel of Figure \ref{fig:evolution}; 
the color code denotes time and is consistent between the panels.
It is straightforward to see that the measured postshock CR spectra (solid lines) are significantly steeper than the standard NLDSA prediction.
They are even  steeper than $p^{-4}$ and match very well the slope calculated using $\trt$, namely
\begin{equation}\label{eq:qtilde}
    \Tilde{q}\equiv \frac{3\trt}{\trt-1}=
    \frac{3\rt}{\rt-1-\alpha}.
\end{equation}
This is plotted as dashed lines in the bottom panel Figure \ref{fig:evolution}.
We have extensively checked this result against simulation parameters such as number of particles per cell, box transverse size, grid resolution, and time step choice; 
we rule out that the steep spectra are a numerical artifact to the best of our knowledge.

\section{Energy Gain vs Escape Probability}
\begin{figure}[t]
\includegraphics[width=.48\textwidth,clip=true,trim= 0 185 0 0]{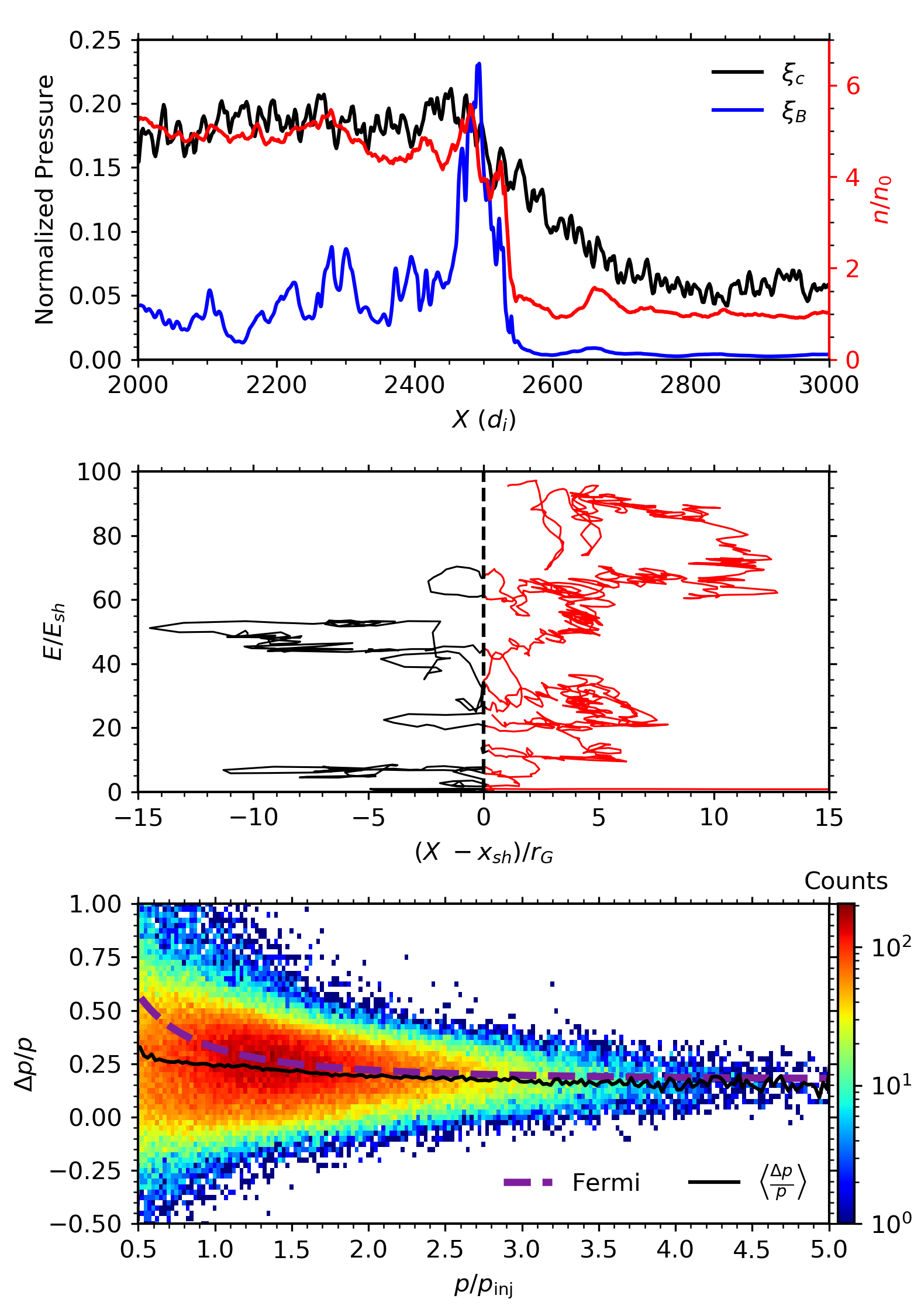}
\caption{Top panel: normalized self-generated CR and magnetic pressure (black and blue/left axis), and number density (red/right axis) averaged over $y$ and plotted along $x$ at $400 \ocii$.
Bottom panel: Energy vs distance from the shock normalized by gyro-radius for a representative CR.
Red (black) segments denote the upstream (downstream) portion of the CR trajectory.
}\label{fig:profile}
\end{figure}

Following the elegant argument of \cite{bell78a}, one can see the CR power-laws arise from a balance between the incremental change in momentum ($\mathcal{G}$) and the probability of escaping the accelerator ($\mathcal{P}$) for each acceleration cycle.
If the CR particle speed $v\gg u$, the average of these quantities over the CR pitch-angle distributions read, respectively:
\begin{equation}\label{eq:gp}
 \mathcal{G} \equiv\left\langle \frac{\Delta p}{p}\right\rangle \simeq \frac{4}{3}\frac{\tu1-\tu2}{v};
 \quad
 \mathcal{P}\simeq \frac{4\tu2}{v},
\end{equation}
where $\Tilde{u}$ represents the bulk flow speed plus the Alfv\'enic drift contribution.
This balance leads to a momentum spectral slope of 
\begin{equation}\label{eq:qbell}
    q \simeq 3 + \frac{\mathcal{P}}{\mathcal{G}} \simeq 
   3+ \frac{3\tu2}{\tu1-\tu2}.
\end{equation}
if $\Tilde{u}= u$ and $u_1=4 u_2$, one recovers the standard $q=4$.
Here we consider the momentum gain, rather than the energy gain, because it pertains to both non-relativistic and relativistic particles; in fact, spectra are power-laws in momentum, not necessarily in energy.
Also, the speeds that matter are those of the reference frame where CR are isotropic, which is why $\Tilde{u}$ appears instead of $u$.

The natural question is then: are CR-modified shocks spectra steeper than the standard prediction because the energy gain is reduced or because the escape probability is increased?
The solutions proposed in the literature argued that the culprit is the smaller $\mathcal{G}$ that arises from the non-negligible correction due to the wave speed in the amplified field, i.e., from the fact that $\tu1=u_1-\w1<u_1$ \citep{zirakashvili+08b,caprioli+10b, caprioli12};
the same type of correction may be also seen as an energy loss for the CRs that do work in order to amplify the magnetic field in the precursor \citep{bell+19}.

Instead, here we show that the main effect is actually due to the postcursor correction, i.e., to the fact that $\tu2=u_2+\w2>u_2$;
such a correction, which is larger in magnitude than the one in the precursor ($\alpha>\alpha_1$), provides both a reduction of $\mathcal{G}$ and an increase in $\mathcal{P}$, but has a greater impact on the escape probability. 
Physically speaking, the CR spectrum becomes steeper because CR are advected away downstream at a greater rate than in unmodified shocks.

\begin{figure}[t]
\includegraphics[width=.48\textwidth,clip=true,trim= 0 0 0 355]{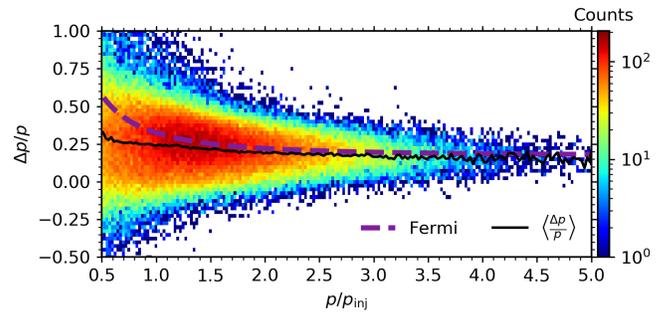}
\caption{
Histograms of the incremental change in momentum as a function of initial momentum for downstream-upstream-downstream Fermi cycles of many tracked CRs.
The standard Fermi prediction (purple dashed lines) is in good agreement with the distribution average above $\pinj$ (black line), implying that steeper spectra are not due to less efficient acceleration or CR losses; instead, they are due to enhanced escape downstream.
}\label{fig:acceleration}
\end{figure}

The top panel of Figure \ref{fig:profile} shows the normalized CR and magnetic pressure (black and blue respectively, left axis) and the density profile (red, right axis) around the shock;
the bottom panel shows the trajectory and energy gain for a tracked CR particle: the red and black regions correspond to the upstream and downstream portion of such a trajectory.
For each tracked particle, we calculate the incremental momentum gain $\Delta p/p$ for each downstream-upstream-downstream, DSA cycle.
The distribution of such momentum gains for many CRs is plotted in Figure \ref{fig:acceleration} and compared with the standard Fermi prediction for $\mathcal{G}$ in Equation \ref{eq:gp}.
The thin black line indicates the average $\mathcal{G}(p)$, which is in good agreement with such a prediction (purple dashed line).
It is also interesting to notice how below the injection momentum $\pinj=\sqrt{10} M v_A$  \citep[chosen following the theory developed in][]{caprioli+15} the distribution of gains broadens and deviates from the Fermi prediction;
this is another independent measurement of the minimum momentum needed for particles to be injected into DSA.

In summary, steeper spectra do not arise because of a reduced energy gain in their Fermi cycles \citep[e.g.,][]{zirakashvili+08b,caprioli12}, or because of the energy that CRs channel into magnetic field amplification \citep{bell+19}, but rather because of an enhanced escape rate in the postcursor, which makes $\mathcal{P}$ larger than the standard value by a factor of $1+\alpha$.   

\section{Downstream evolution}\label{sec:dse}
\begin{figure}[t]
\includegraphics[width=.48\textwidth,clip=true,trim= 0 0 0 0]{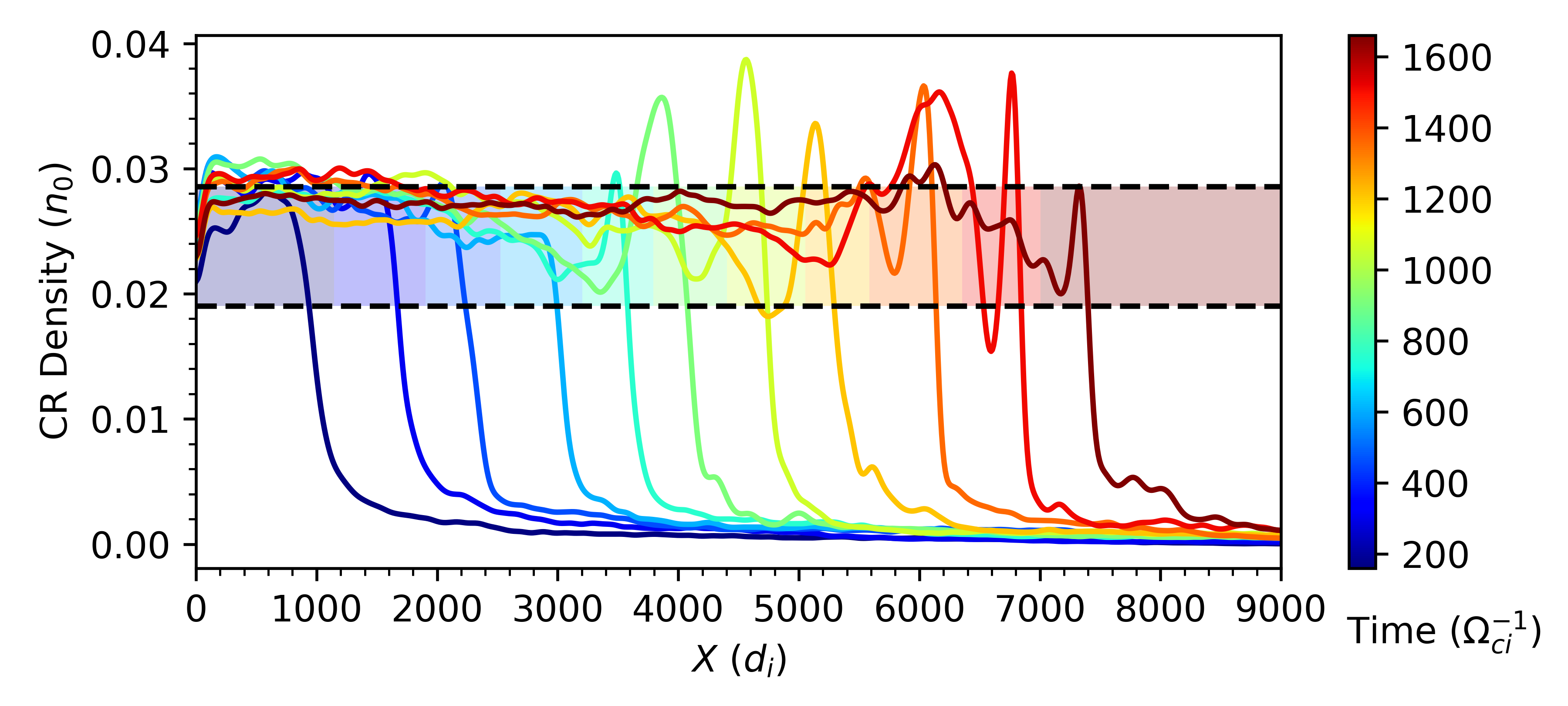}
\caption{Time evolution of the CR density profile. Colored band corresponds to a range between 66.7\% and 100\% of the average far downstream density, color coded for time.
Note the upstream precursor, an overshoot at the shock, and a gradual compression in the postcursor (see Section~\ref{sec:dse} for details).
}\label{fig:rho}
\end{figure}

In the standard theory of NLDSA, the downstream is usually assumed to be homogeneous, a hypothesis that may be violated if the velocity of the scattering centers varied behind the shock.
In general, this would lead to both a fluid speed profile and spatially-varying CR distribution.

To quantify this effect, we consider the 1D, stationary transport equation for the isotropic part of the CR distribution function, $f(x,p)$ \citep{skilling75a}
\begin{equation}\label{eq:parker}
    \Tilde{u}(x)\frac{\partial f}{\partial x} =
    \frac{\partial}{\partial x}\left[D(x,p)\frac{\partial f}{\partial x}\right]+\frac{p}{3}\frac{d \Tilde{u}}{dx}\frac{\partial f}{\partial p},
\end{equation}
where the CR advection speed $\Tilde{u}=u+v_w$ should tend to the fluid speed $u_2$ beyond some distance $L$ in the downstream.
Physically speaking, $L$ may correspond to the length-scale over which the self-generated magnetic turbulence is damped, for instance via non-linear Landau damping \citep{volk+81, lee+73} or other phenomena typical of high-$\beta$ plasmas \citep{squire+17, kunz+20}.
In fact, if in the precursor there is quasi equipartition between magnetic fields and thermal energy (\S3 of Paper I), for strong shocks the thermal pressure jump ($\propto M^2$) is larger than the magnetic one ($\propto \rs^2$), returning a large $\beta\approx M^2/16$.
We stress that it is necessary for the self-generated magnetic field to ``unwind" at some point behind the shock, since conservation laws enforce the asymptotic configuration far downstream to match the one far upstream, at least for $B_x$.
Nevertheless, the spatial extent of the post-shock region with enhanced fields is easily much larger than the CR diffusion scales \citep[e.g.,][]{morlino+10,morlino+12, ressler+14}.

The slope of the CR distribution function is simply 
\begin{equation}
    q= -\frac{p}{f}\frac{\partial f}{\partial p}
\end{equation}
and Equation \ref{eq:parker} can be solved with standard techniques, i.e., by integrating across the shock as well as from the shock to upstream infinity, where $f\to 0$. \citep[e.g.,][]{blasi02,amato+06,caprioli+10b}.
For instance, neglecting the precursor, in the limit $D\to 0$ (pure advection), behind the shock one gets
\begin{equation}
    \frac{\partial f}{\partial x}\simeq -
    \frac{q}{3}\frac{d \Tilde{u}}{dx}\frac{f}{\Tilde{u}}, 
\end{equation}
which describes adiabatic compression and reduces to $f\to$ constant if $d \Tilde{u}/{dx}\to 0$: as the CR drift velocity approaches zero, the CR population increases its density without any spectral change. 

Restoring diffusion, with the simplifying assumption that $u_w$ decreases linearly over the postcursor on a length-scale $L$, i.e.,:
\begin{equation}
    \frac{d \Tilde{u}}{dx} \simeq  - \frac{v_w}{u_2 + v_w}
    \frac{\Tilde{u}}{L} = - \frac{\alpha}{\alpha+1} \frac{\Tilde{u}}{L},
\end{equation}
Equation \ref{eq:parker} can be promptly solved and leads to:
\begin{equation}
    q = \frac{3 r}{ r-1- \alpha + \lambda(p)}; 
    \quad 
    \lambda(p)\equiv \frac{\alpha}{\alpha-1}\frac{D_2(p)}{u_2 L}.
\end{equation}
Since $\lambda>0$, the vanishing of the CR drift over the postcursor induces a \emph{hardening} of the spectrum; however, for all the momenta for which the CR diffusion length $D(p)/u_2\ll L$ and $\lambda(p)\ll 1$ the effect is negligible.
Again, such a hardening comes from the differential adiabatic compression of CRs with different momentum, with larger-$p$ particles experiencing more compression.
For the hardening to be global, $L$ should be smaller than the region where diffusion is enhanced by the self-generated magnetic field, at odds with the very nature of the postcursor. 

Even if the effect on the spectrum is negligible, it is easy to estimate (always in the limit $L\gg D/\Tilde{u}$) the adiabatic compression of CR in the postcursor as
\begin{equation}
    \frac{\Delta f}{f} \simeq \frac{q}{3 (\alpha+1)},
\end{equation}
i.e., the CR distribution function should increase of $\mathcal O (1)$ for $q\approx 4$ and $\alpha\approx 0.5$.
Figure \ref{fig:rho} shows the average CR density profile as a function of time (color coded) for our benchmark run; 
three features can be noticed: 
1) an upstream exponential profile, which corresponds to the classical CR precursor;
2) an overshoot at the shock, where the density exceeds the asymptotic one, $\rt$, which has a quasi-periodic nature, as discussed in Paper I;
3) a quite gradual rise in the downstream, on the postcursor extent, which is the result of the effect just discussed.
It is important to stress that the spectrum at the shock is only affected by what happens within one diffusion length $D(p)/u_2$ downstream, while such an extra compression occurs at the end of the postcursor; 
this must be reckoned with when investigating the origin of either synchrotron emission, which should track the magnetized postcursor region, or hadronic and bremsstrahlung emissions, which track plasma and CR density.

\subsection{Dependence on Mach Number}
\begin{figure}[t]
\includegraphics[width=0.48\textwidth, clip=true,trim= 0 0 0 0]{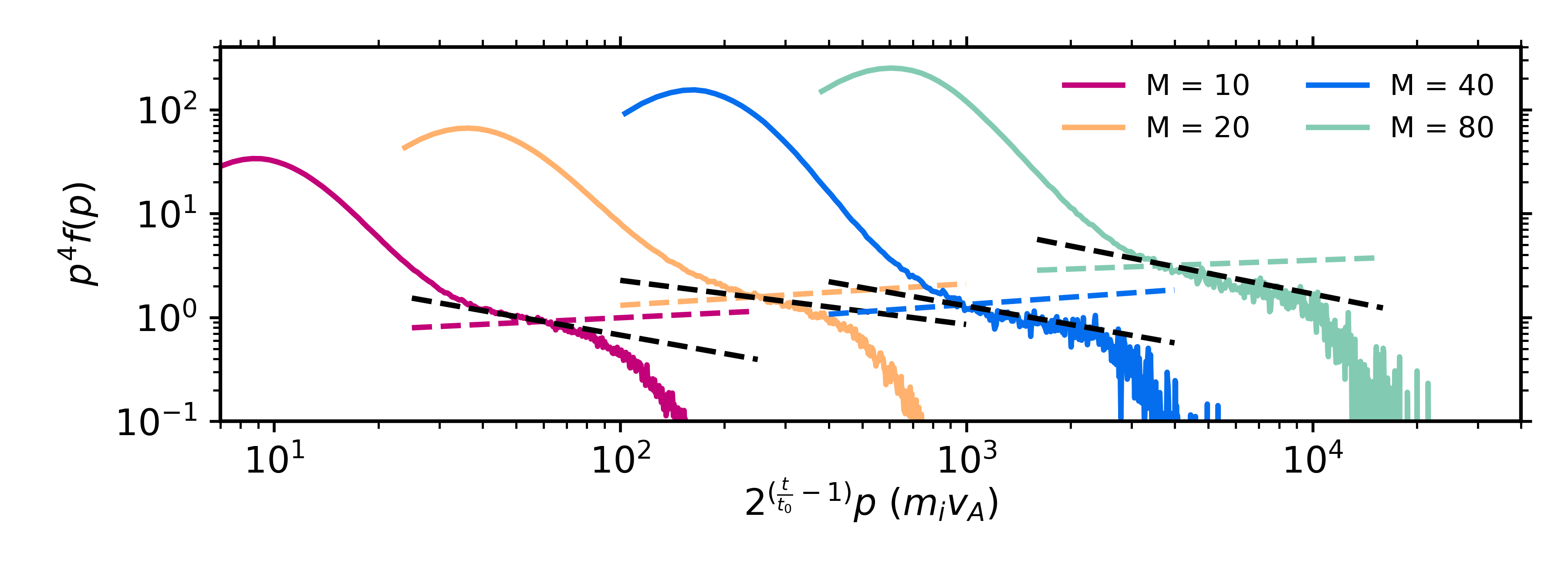}
\caption{Post-shock particle spectra for different Mach number simulations (see Table \ref{tab:sims}), at $t\approx 370\ocii$. The standard DSA prediction along with the modified prediction presented in this paper are shown as the dashed colored and black lines respectively.
}\label{fig:machs}
\end{figure}
Figure \ref{fig:machs} shows the postshock spectrum for shocks of different Mach numbers $M=10,20,40,80$, for the same simulations discussed in Paper I. 
Also, in this case we show both the standard (flatter, color) and revised (steeper, black) predictions and it is clear that the latter (Equation \ref{eq:qtilde}) is consistent with the simulations. 
Details about the these simulations and the measured values to predict both slopes are presented in Table~\ref{tab:sims}.
\begin{center}
\begin{table}
\begin{tabular}{|ccccc|}
        \hline
        $M$ & $\xi_c$ & $\xi_B$ & $\rt$ & $\Tilde{q}-q_{\rm DSA}$\\
        \hline
    10 & 0.072 & 0.035 & 4.54 & 0.75\\
    20 & 0.099 & 0.028 & 4.54 & 0.63\\
    40 & 0.102 & 0.033 & 4.61 & 0.82\\
    80 & 0.100 & 0.018 & 4.37 & 0.78\\
         \hline
\end{tabular}
\caption{Physical parameter for runs with different Mach numbers. From left to right: Mach number ($M$), normalized CR pressure ($\xi_c$), normalized magnetic pressure ($\xi_B$), total compression ratio ($\rt$), difference between the revised and the standard DSA momentum slope ($\Tilde{q}-q_{\rm DSA}$). The revised (standard) slopes are shown in Figure~\ref{fig:machs} as dashed black (color) lines.}
\label{tab:sims}
\end{table}
\end{center}

The theory for the hydrodynamic modifications discussed in Paper I as well as the theory for the spectrum of the accelerated particles presented here both rely on assumptions about the nature of the self-generated magnetic turbulence. Arguably the most crucial of these assumptions is that the CRs drift away from the shock at the local Alfv\'en speed; an assumption which is inspired by, but also validated with, self-consistent simulations.
This is especially relevant for very strong shocks, where the instability that drives magnetic field amplification should be in the Bell regime ($M\gtrsim 30$), for which the intuition based on the quasi-linear theory may stumble. 
Exploring even stronger shocks ($M\gtrsim 100$) with hybrid techniques is computationally prohibitive, however, it is reasonable that any poorly-magnetized shock, such that the upstream magnetic field amplification is driven by  the Weibel, that is able to inject protons, may quickly (on few growth times, typically corresponding to $10-100\ocii$) transition to a shock with Alfv\'enic Mach numbers comparable to those studied in this work.   

On the other hand, we stress that all of the effects of the postcursor should vanish in cases where magnetic field amplification is not prominent ($\delta B/B_0\lesssim 1$), such as very oblique shocks that do not spontaneously inject particles into DSA or at weak shocks with small Alfv\'enic Mach number \citep{caprioli+14a,caprioli+14b}, regimes which we will cover in a future work.
However, in the presence of energetic seed particles, which may be injected even at oblique shocks \citep{caprioli+18}, a postcursor may still be generated, provided that re-accelerated particles can drive sufficiently strong magnetic turbulence.

\section{Phenomenological Implications\label{sec:phenom}}
The main implication of these findings is that CR-modified shocks naturally lead to CR spectra steeper than the standard NLDSA prediction, and generally steeper than $p^{-4}$, which should account for the phenomenology of $\gamma$-ray bright SNRs, radio SNe, and even of Galactic CRs, as discussed in Section \ref{sec:intro}.

Given that CR modifications produce spectra steeper than $E^{-2}$, it is unlikely that concave spectra would develop as predicted by the classical NLDSA theory. 
In fact, if most of the energy in the CR distribution is in trans-relativistic particles, the extent of the precursor is ultimately set by $\sim $ GeV protons, and all of the relativistic particles must experience the same precursor/postcursor and thus undergo DSA ``feeling'' the same total compression ratio, $\rt$.
There is still room for a further steepening at non-relativistic energies (provided that $q\lesssim 5$, otherwise the precursor would be set by particles with $\pinj$), which may have a radiative signature in the synchrotron emission in radio SNe, where strong magnetic fields, $\lesssim 1$G, may reveal emission from sub-GeV electrons.  

While a detailed modeling of the different environments is left to forthcoming publications, it is worth pointing out a general feature of the steepening induced by the presence of the postcursor. 
Since the CR spectral slope eventually depends on $\rt$ and $\alpha$ (Equation \ref{eq:qtilde}), a multi-wavelength study of an astrophysical object can constrain shock speed and post-shock magnetic field via the modified jump conditions outlined in Paper I, and vice-versa.
For instance, in SNRs, the post-shock magnetic field can be constrained via synchrotron emission \citep[thickness of X-ray rims, radio/X-ray flux, see, e.g., the review][and references therein]{caprioli15p}, while the density and shock speed can be estimated via high-resolution X-ray observations \citep[e.g.,][]{warren+05,gamil+08, miceli+12}; 
the slope of the synchrotron spectrum should then be consistent with the presented theory, which offers a redundant way of testing our prescription using multi-wavelength observations. 

In general
\begin{equation}
    \alpha \simeq 5\times 10^{-3}
    \frac{B_2}{\mu {\rm G}} 
    \frac{1000 {\rm km~s}^{-1}}{\vsh}
    \left(\frac{\rt}{5}
    \frac{{\rm cm}^{-3}}{n_0}\right)^{\frac{1}{2}}
\end{equation}

Different analyses of Tycho's SNR broadband emission  \citep[e.g.,][]{morlino+12,slane+14} converge on showing that the radio to $\gamma$-ray spectrum can be fitted with a shock speed of $\vsh\sim 5000$km s$^{-1}$, a preshock number density of $n_0 \approx \rho_0/m_p \sim 0.3$ cm$^{-3}$, and a postshock magnetic field of $B_2\sim 300\mu$G, from which one would infer $\alpha \approx 0.55$ and in turn $q\approx 4.3$, in very good agreement with the slope inferred from both radio and $\gamma$-rays, which are a measurement of (uncooled) electron and proton spectra.

In a quite different regime, for a typical Type Ib/c radio SN  \citep[e.g., SN1983N, see table 1 in][]{chevalier+06}, where $\vsh\sim 42000$km s$^{-1}$, $n_0\sim 5600$ cm$^{-3}$, and $B_2\sim 0.56$G, one would get $\alpha\approx 0.96$ and eventually $q\approx 5$, in remarkable agreement with the fact that the radio emission from typical extragalactic SNe requires very steep spectra, $\propto E^{-3}$.
Discussing further examples goes beyond the scope of this paper, but it is noteworthy that our theory seems to naturally capture the steepness of both SN and SNR spectra for reference objects. 

With a sightly different twist, we also note that, if shock speed and density are known, one could use the presented framework to estimate the value of the downstream magnetic field directly from the slope of the radio spectrum, which is straightforward to measure in the non-self-absorbed regime. 
In other words, it should be possible to remove one free parameter in the fitting of the non-thermal emission from radio SNe, kilonovae, GRB afterglows, and so on. 
We acknowledge that the uncertainties in the estimate of environmental parameters in astrophysical contexts may be quite large, so that the feasibility of this analysis must be assessed on a case-by-case basis. 

\section{Conclusion}
In this work we use self-consistent hybrid simulations to study the effects of self-generated CRs and subsequent turbulent magnetic fields on the spectral index of the CR power-law distribution.
In a companion paper (\cite{haggerty+20}, reffered to as Paper I in this work), we showed that the CRs and associated magnetic turbulence modify the hydrodynamics of the shock, leading to a shock compression ratio larger than in the standard theory;
in this work we show that, while the standard DSA theory  predicts that a larger compression induces a flattening in the CR spectra, in reality CR-modified shocks lead to particle spectra \emph{steeper} than the test-particle prediction of $p^{-4}$, or $E^{-2}$ for relativistic CRs.

We present a revised theory of non-linear DSA (NLDSA) which accounts for the newly identified postcursor region, a region comprised of amplified magnetic fields just downstream of the shock, discussed in detail in Paper I.
The standard DSA prediction is modified because CRs are isotropized in the postcursor frame, a frame that moves away from the shock with the local Alfv\'en speed relative to the downstream bulk flow.
This enhanced CR advection in the downstream region increases the rate at which CRs escape from the shock, resulting in a spectrum that is not only steeper than the NLDSA prediction, but steeper than even the standard DSA prediction of $q = 4$ in momentum space (Figure \ref{fig:evolution}).
Such steep spectra cannot be ascribed to a reduced rate of acceleration, or to losses due to the energy that goes into magnetic field generation (which agree with the standard Fermi theory, Figure \ref{fig:acceleration}), but are rather due to the enhanced CR escape downstream.  

The presented theory is inspired and validated by hybrid simulation of parallel shocks over a range of different Mach numbers, from relatively modest ($M = 10$) to strong ($M=80$) shocks (Figure \ref{fig:machs}). 
Simulations consistently show enhanced particle escape downstream, and correspondingly steeper CR spectra in agreement with the theory put forth in this work.
A complete theory for the saturation of the magnetic field amplification in very strong shocks, in which the CR spectrum extends for many decades into the relativistic regime and the Bell instability plays a dominant role, is still missing;
moreover, additional effects such as the acoustic instability \citep{drury-falle86} or the turbulent amplification triggered by the shock corrugation \citep{richtmyer60,meshkov72, caprioli+13} may contribute to enhance the post-shock magnetic field \citep[also see][]{giacalone+07, inoue+09,yokoyama+20}.
Therefore, we cannot predict from first principles the exact amount of steepening induced by the presence of a postcursor, but the typical values inferred in astrophysical situations suggest that the effect must be taken into account (\S\ref{sec:phenom}).

We considered the phenomenological implications of the steeper spectra for astrophysical shocks, which suggests that many modern observations of $\gamma$-ray bright SNRs, radio SNe and even Galactic CRs more generally are in better agreement with these findings.
We also discussed an intriguing corollary of these results, namely, how measurements of astrophysical shock compression ratios and spectral index can be used to infer post-shock magnetic field strength and vice-versa.
This is outlined for paradigmatic SN and SNR values and found to be in agreement with observations.

These results, along with those in paper I, present clear and detailed evidence of the existence and relevance of CR-modified shocks, simultaneously resolving a growing tension between the standard DSA theory and observations, further reinforcing the prominence of collisionless shocks as efficient accelerators of non-thermal particles.

\software{\dHybridR~\citep{haggerty+19a}}

\acknowledgments
We would like to thank  E.~Amato, A.~Spitkovsky, D.~Eichler, L.~O'C. Drury, S.~Schwartz, and L.~Wilson III for stimulating and constructive discussions. 
This research was partially supported by NASA (grant NNX17AG30G, 80NSSC18K1218, and 80NSSC18K1726), NSF (grants AST-1714658, AST-1909778, PHY-1748958, PHY-2010240), and by the International Space Science Institute’s (ISSI) International Teams program.
Simulations were performed on computational resources provided by the University of Chicago Research Computing Center, the NASA High-End Computing Program through the NASA Advanced Supercomputing Division at Ames Research Center, and XSEDE TACC (TG-AST180008).

\bibliography{Total}

\begin{thebibliography}{}
\expandafter\ifx\csname natexlab\endcsname\relax\def\natexlab#1{#1}\fi

\bibitem[{{Aguilar et al. [AMS Collaboration]}(2016)}]{ams16b}
{Aguilar et al. [AMS Collaboration]}, M. 2016, Phys. Rev. Lett., 117, 231102

\bibitem[{{Amato} \& {Blasi}(2006)}]{amato+06}
{Amato}, E., \& {Blasi}, P. 2006, MNRAS, 371, 1251

\bibitem[{{Axford} {et~al.}(1978){Axford}, {Leer}, \& {Skadron}}]{axford+78}
{Axford}, W.~I., {Leer}, E., \& {Skadron}, G. 1978, in International Cosmic Ray
  Conference, Vol.~11, ICRC, 132--137

\bibitem[{{Bell}(1978)}]{bell78a}
{Bell}, A.~R. 1978, MNRAS, 182, 147

\bibitem[{{Bell} {et~al.}(2019){Bell}, {Matthews}, \& {Blundell}}]{bell+19}
{Bell}, A.~R., {Matthews}, J.~H., \& {Blundell}, K.~M. 2019, \mnras, 488, 2466

\bibitem[{Bell {et~al.}(2011)Bell, Schure, \& Reville}]{bell+11}
Bell, A.~R., Schure, K.~M., \& Reville, B. 2011, \mnras, 418, 1208

\bibitem[{{Blandford} \& {Ostriker}(1978)}]{blandford+78}
{Blandford}, R.~D., \& {Ostriker}, J.~P. 1978, ApJL, 221, L29

\bibitem[{{Blasi}(2002)}]{blasi02}
{Blasi}, P. 2002, APh, 16, 429

\bibitem[{{Blasi} \& {Amato}(2012{\natexlab{a}})}]{blasi+11a}
{Blasi}, P., \& {Amato}, E. 2012{\natexlab{a}}, \jcap, 1, 10

\bibitem[{{Blasi} \& {Amato}(2012{\natexlab{b}})}]{blasi+11b}
---. 2012{\natexlab{b}}, \jcap, 1, 11

\bibitem[{Blasi {et~al.}(2007)Blasi, Amato, \& Caprioli}]{blasi+07}
Blasi, P., Amato, E., \& Caprioli, D. 2007, Monthly Notices of the Royal
  Astronomical Society, 375, 1471

\bibitem[{{Blasi} {et~al.}(2012){Blasi}, {Morlino}, {Bandiera}, {Amato}, \&
  {Caprioli}}]{blasi+12a}
{Blasi}, P., {Morlino}, G., {Bandiera}, R., {Amato}, E., \& {Caprioli}, D.
  2012, \apj, 755, 121

\bibitem[{{Caprioli}(2011)}]{caprioli11}
{Caprioli}, D. 2011, \jcap, 5, 26

\bibitem[{{Caprioli}(2012)}]{caprioli12}
---. 2012, \jcap, 7, 38

\bibitem[{Caprioli(2015)}]{caprioli15p}
Caprioli, D. 2015, in International Cosmic Ray Conference, Vol.~34, 34th
  International Cosmic Ray Conference (ICRC2015), ed. A.~S. {Borisov}, V.~G.
  {Denisova}, Z.~M. {Guseva}, E.~A. {Kanevskaya}, M.~G. {Kogan}, A.~E.
  {Morozov}, V.~S. {Puchkov}, S.~E. {Pyatovsky}, G.~P. {Shoziyoev}, M.~D.
  {Smirnova}, A.~V. {Vargasov}, V.~I. {Galkin}, S.~I. {Nazarov}, \& R.~A.
  {Mukhamedshin}, 8

\bibitem[{{Caprioli} {et~al.}(2010){Caprioli}, {Amato}, \&
  {Blasi}}]{caprioli+10b}
{Caprioli}, D., {Amato}, E., \& {Blasi}, P. 2010, APh, 33, 307

\bibitem[{{Caprioli} {et~al.}(2009{\natexlab{a}}){Caprioli}, {Blasi}, \&
  {Amato}}]{caprioli+09b}
{Caprioli}, D., {Blasi}, P., \& {Amato}, E. 2009{\natexlab{a}}, MNRAS, 396,
  2065

\bibitem[{{Caprioli} {et~al.}(2008){Caprioli}, {Blasi}, {Amato}, \&
  {Vietri}}]{caprioli+08}
{Caprioli}, D., {Blasi}, P., {Amato}, E., \& {Vietri}, M. 2008, ApJ Lett, 679,
  L139

\bibitem[{{Caprioli} {et~al.}(2009{\natexlab{b}}){Caprioli}, {Blasi}, {Amato},
  \& {Vietri}}]{caprioli+09a}
---. 2009{\natexlab{b}}, MNRAS, 395, 895

\bibitem[{{Caprioli} \& {Haggerty}(2019)}]{caprioli+19p}
{Caprioli}, D., \& {Haggerty}, C. 2019, in International Cosmic Ray Conference,
  Vol.~36, 36th International Cosmic Ray Conference (ICRC2019), 209

\bibitem[{{Caprioli} {et~al.}(2015){Caprioli}, {Pop}, \&
  {Spitkovsky}}]{caprioli+15}
{Caprioli}, D., {Pop}, A., \& {Spitkovsky}, A. 2015, ApJ Letters, 798, 28

\bibitem[{{Caprioli} \& {Spitkovsky}(2013)}]{caprioli+13}
{Caprioli}, D., \& {Spitkovsky}, A. 2013, \apjl, 765, L20

\bibitem[{{Caprioli} \& {Spitkovsky}(2014{\natexlab{a}})}]{caprioli+14a}
---. 2014{\natexlab{a}}, \apj, 783, 91

\bibitem[{{Caprioli} \& {Spitkovsky}(2014{\natexlab{b}})}]{caprioli+14b}
---. 2014{\natexlab{b}}, \apj, 794, 46

\bibitem[{{Caprioli} {et~al.}(2017){Caprioli}, {Yi}, \&
  {Spitkovsky}}]{caprioli+17}
{Caprioli}, D., {Yi}, D.~T., \& {Spitkovsky}, A. 2017, \prl, 119, 171101

\bibitem[{Caprioli {et~al.}(2018)Caprioli, Zhang, \& Spitkovsky}]{caprioli+18}
Caprioli, D., Zhang, H., \& Spitkovsky, A. 2018, JPP, arXiv:1801.01510

\bibitem[{{Cassam-Chena{\"\i}} {et~al.}(2008){Cassam-Chena{\"\i}}, {Hughes},
  {Reynoso}, {Badenes}, \& {Moffett}}]{gamil+08}
{Cassam-Chena{\"\i}}, G., {Hughes}, J.~P., {Reynoso}, E.~M., {Badenes}, C., \&
  {Moffett}, D. 2008, \apj, 680, 1180

\bibitem[{Chevalier \& Fransson(2006)}]{chevalier+06}
Chevalier, R.~A., \& Fransson, C. 2006, \apj, 651, 381

\bibitem[{{Ellison} {et~al.}(2000){Ellison}, {Berezhko}, \&
  {Baring}}]{ellison+00}
{Ellison}, D.~C., {Berezhko}, E.~G., \& {Baring}, M.~G. 2000, \apj, 540, 292

\bibitem[{{Ellison} \& {Reynolds}(1991)}]{ellison+91}
{Ellison}, D.~C., \& {Reynolds}, S.~P. 1991, \apj, 382, 242

\bibitem[{{Evoli} {et~al.}(2019{\natexlab{a}}){Evoli}, {Aloisio}, \&
  {Blasi}}]{evoli+19a}
{Evoli}, C., {Aloisio}, R., \& {Blasi}, P. 2019{\natexlab{a}}, \prd, 99, 103023

\bibitem[{{Evoli} {et~al.}(2019{\natexlab{b}}){Evoli}, {Morlino}, {Blasi}, \&
  {Aloisio}}]{evoli+19b}
{Evoli}, C., {Morlino}, G., {Blasi}, P., \& {Aloisio}, R. 2019{\natexlab{b}},
  arXiv e-prints, arXiv:1910.04113

\bibitem[{{Gargat{\'e}} {et~al.}(2007){Gargat{\'e}}, {Bingham}, {Fonseca}, \&
  {Silva}}]{gargate+07}
{Gargat{\'e}}, L., {Bingham}, R., {Fonseca}, R.~A., \& {Silva}, L.~O. 2007,
  Comp. Phys. Commun., 176, 419

\bibitem[{{Giacalone} \& {Jokipii}(2007)}]{giacalone+07}
{Giacalone}, J., \& {Jokipii}, J.~R. 2007, \apjl, 663, L41

\bibitem[{{Haggerty} \& {Caprioli}(2019)}]{haggerty+19a}
{Haggerty}, C.~C., \& {Caprioli}, D. 2019, \apj, 887, 165

\bibitem[{{Haggerty} \& {Caprioli}(2020)}]{haggerty+20}
---. 2020, Submitted to \apj, arXiv:2008.12308

\bibitem[{{Hanusch} {et~al.}(2019){Hanusch}, {Liseykina}, {Malkov}, \&
  {Aharonian}}]{hanusch+19}
{Hanusch}, A., {Liseykina}, T.~V., {Malkov}, M., \& {Aharonian}, F. 2019, \apj,
  885, 11

\bibitem[{{Inoue} {et~al.}(2009){Inoue}, {Yamazaki}, \& {Inutsuka}}]{inoue+09}
{Inoue}, T., {Yamazaki}, R., \& {Inutsuka}, S.-i. 2009, \apj, 695, 825

\bibitem[{{Jones} \& {Ellison}(1991)}]{jones+91}
{Jones}, F.~C., \& {Ellison}, D.~C. 1991, Space Science Reviews, 58, 259

\bibitem[{{Kang} {et~al.}(2013){Kang}, {Jones}, \& {Edmon}}]{kang+13}
{Kang}, H., {Jones}, T.~W., \& {Edmon}, P.~P. 2013, \apj, 777, 25

\bibitem[{{Kang} \& {Ryu}(2018)}]{kang+18}
{Kang}, H., \& {Ryu}, D. 2018, \apj, 856, 33

\bibitem[{{Kirk} {et~al.}(1996){Kirk}, {Duffy}, \& {Gallant}}]{kirk+96}
{Kirk}, J.~G., {Duffy}, P., \& {Gallant}, Y.~A. 1996, \aap, 314, 1010

\bibitem[{{Krymskii}(1977)}]{krymskii77}
{Krymskii}, G.~F. 1977, Akademiia Nauk SSSR Doklady, 234, 1306

\bibitem[{{Kunz} {et~al.}(2020){Kunz}, {Squire}, {Schekochihin}, \&
  {Quataert}}]{kunz+20}
{Kunz}, M.~W., {Squire}, J., {Schekochihin}, A.~A., \& {Quataert}, E. 2020,
  arXiv e-prints, arXiv:2006.08940

\bibitem[{{Lee} \& {V{\"o}lk}(1973)}]{lee+73}
{Lee}, M.~A., \& {V{\"o}lk}, H.~J. 1973, \apss, 24, 31

\bibitem[{{Malkov} \& {Aharonian}(2019)}]{malkov+19}
{Malkov}, M.~A., \& {Aharonian}, F.~A. 2019, \apj, 881, 2

\bibitem[{Malkov {et~al.}(2012)Malkov, Diamond, \& Sagdeev}]{malkov+12}
Malkov, M.~A., Diamond, P.~H., \& Sagdeev, R.~Z. 2012, Physical Review Letters,
  108, 081104

\bibitem[{{Malkov} \& {O'C. Drury}(2001)}]{malkov+01}
{Malkov}, M.~A., \& {O'C. Drury}, L. 2001, Rep. Prog. Phys., 64, 429

\bibitem[{{Meshkov}(1972)}]{meshkov72}
{Meshkov}, E.~E. 1972, Fluid Dynamics, 4, 101

\bibitem[{{Miceli} {et~al.}(2012){Miceli}, {Bocchino}, {Decourchelle},
  {Maurin}, {Vink}, {Orlando}, {Reale}, \& {Broersen}}]{miceli+12}
{Miceli}, M., {Bocchino}, F., {Decourchelle}, A., {et~al.} 2012, \aap, 546, A66

\bibitem[{{Morlino} {et~al.}(2010){Morlino}, {Amato}, {Blasi}, \&
  {Caprioli}}]{morlino+10}
{Morlino}, G., {Amato}, E., {Blasi}, P., \& {Caprioli}, D. 2010, \mnras, 405,
  L21

\bibitem[{{Morlino} {et~al.}(2012){Morlino}, {Bandiera}, {Blasi}, \&
  {Amato}}]{morlino+12b}
{Morlino}, G., {Bandiera}, R., {Blasi}, P., \& {Amato}, E. 2012, \apj, 760, 137

\bibitem[{{Morlino} \& {Blasi}(2016)}]{morlino+16}
{Morlino}, G., \& {Blasi}, P. 2016, \aap, 589, A7

\bibitem[{{Morlino} {et~al.}(2013){Morlino}, {Blasi}, {Bandiera}, {Amato}, \&
  {Caprioli}}]{morlino+13}
{Morlino}, G., {Blasi}, P., {Bandiera}, R., {Amato}, E., \& {Caprioli}, D.
  2013, \apj, 768, 148

\bibitem[{{Morlino} \& {Caprioli}(2012)}]{morlino+12}
{Morlino}, G., \& {Caprioli}, D. 2012, A\&A, 538, A81

\bibitem[{{O'C. Drury}(1983)}]{drury83}
{O'C. Drury}, L. 1983, Reports of Progress in Physics, 46, 973

\bibitem[{{O'C. Drury} \& {Falle}(1986)}]{drury-falle86}
{O'C. Drury}, L., \& {Falle}, S.~A.~E.~G. 1986, MNRAS, 223, 353

\bibitem[{{O'C. Drury} \& {V{\"o}lk}(1981)}]{drury-volk81a}
{O'C. Drury}, L., \& {V{\"o}lk}, H.~J. 1981, Ap. J., 248, 344

\bibitem[{{Ressler et al.}(2014)}]{ressler+14}
{Ressler et al.}, S.~M. 2014, \apj, 790, 85

\bibitem[{Richtmyer(1960)}]{richtmyer60}
Richtmyer, R.~D. 1960, Communications on Pure and Applied Mathematics, 13, 297

\bibitem[{{Skilling}(1975)}]{skilling75a}
{Skilling}, J. 1975, MNRAS, 172, 557

\bibitem[{Slane {et~al.}(2014)Slane, Lee, Ellison, Patnaude, Hughes, Eriksen,
  Castro, \& Nagataki}]{slane+14}
Slane, P., Lee, S.-H., Ellison, D.~C., {et~al.} 2014, \apj, 783, 33

\bibitem[{Squire {et~al.}(2017)Squire, Kunz, Quataert, \&
  Schekochihin}]{squire+17}
Squire, J., Kunz, M.~W., Quataert, E., \& Schekochihin, A.~A. 2017, PRL, 119,
  155101

\bibitem[{{Tatischeff}(2009)}]{tatischeff09}
{Tatischeff}, V. 2009, \aap, 499, 191

\bibitem[{{Vladimirov} {et~al.}(2006){Vladimirov}, {Ellison}, \&
  {Bykov}}]{vladimirov+06}
{Vladimirov}, A., {Ellison}, D.~C., \& {Bykov}, A. 2006, \apj, 652, 1246

\bibitem[{{V{\"o}lk} \& {McKenzie}(1981)}]{volk+81}
{V{\"o}lk}, H.~J., \& {McKenzie}, F.~J. 1981, in International Cosmic Ray
  Conference, Vol.~9, {\emph{Characteristics of Cosmic-Ray Shocks in the
  Presence of Wave Dissipation}}, 246--+

\bibitem[{{Warren et al.}(2005)}]{warren+05}
{Warren et al.}, J.~S. 2005, Ap. J., 634, 376

\bibitem[{{Yokoyama} \& {Ohira}(2020)}]{yokoyama+20}
{Yokoyama}, S.~L., \& {Ohira}, Y. 2020, \apj, 897, 50

\bibitem[{{Zirakashvili} \& {Ptuskin}(2008)}]{zirakashvili+08b}
{Zirakashvili}, V.~N., \& {Ptuskin}, V.~S. 2008, astro-ph/0807.2754,
  arXiv:0807.2754

\end{thebibliography}

\end{document}